\documentclass[osajnl,twocolumn,showpacs,superscriptaddress,10pt]{revtex4-1}

\usepackage{amsmath,amssymb,graphicx}
\usepackage{hyperref}      
\usepackage{braket}
\usepackage{relsize}
\usepackage{dsfont} 
\newcommand{\tr}{\mathsmaller T} 
\newcommand{\abs}[1]{\left| #1 \right|} 
\newcommand{\dash}{\textemdash~}
  \hypersetup{
    colorlinks=true,  
    linkcolor=blue,   
    citecolor=blue,  
    urlcolor =blue        
  }

\begin{document}

\title{Modeling quasi-dark states with \\ Temporal Coupled-Mode Theory}
\author{Mario C. M. M. Souza}
\author{Guilherme F. M. Rezende}
\address{``Gleb Wataghin" Physics Institute, University of Campinas, 13083-970 Campinas, SP, Brazil}
\author{Luis A. M. Barea}
\address{Dept. of Electrical Engineering, UFSCAR, 13565-905 Sao Carlos, SP, Brazil}
\author{Gustavo S. Wiederhecker}
\author{Newton C. Frateschi}\email{fratesch@ifi.unicamp.br}
\address{``Gleb Wataghin" Physics Institute, University of Campinas, 13083-970 Campinas, SP, Brazil}

\begin{abstract}
Coupled resonators are commonly used to achieve tailored spectral responses and allow novel functionalities in a broad range of applications, from optical modulation and filtering in integrated photonic circuits to the study of nonlinear dynamics in arrays of resonators.
The Temporal Coupled-Mode Theory (TCMT) provides a simple and general tool that is widely used to model these devices and has proved to yield very good results in many different systems of low-loss, weakly coupled resonators.
Relying on TCMT to model coupled resonators might however be misleading in some circumstances due to the lumped-element nature of the model.
In this article, we report an important limitation of TCMT related to the prediction of dark states.
Studying a coupled system composed of three microring resonators, we demonstrate that TCMT predicts the existence of a dark state that is in disagreement with experimental observations and with the more general results obtained with the Transfer Matrix Method (TMM) and the Finite-Difference Time-Domain (FDTD) simulations.
We identify the limitation in the TCMT model to be related to the mechanism of excitation/decay of the supermodes and we propose a correction that effectively reconciles the model with expected results.
A comparison with TMM and FDTD allows to verify both steady-state and transient solutions of the modified-TCMT model.
The proposed correction is derived from general considerations, energy conservation and the non-resonant power circulating in the system, therefore it provides good insight on how the TCMT model should be modified to eventually account for the same limitation in a different coupled-resonator design.
Moreover, our discussion based on coupled microring resonators can be useful for other electromagnetic resonant systems due to the generality and far-reach of the TCMT formalism.
\end{abstract}

\maketitle 


\section{Introduction}
\label{sec:intro}

Resonant structures are ubiquitous in nanophotonics \cite{Yu2014e,Muller2014,Xu2009b,Souza2014c,Souza2015e,Zhang2015a,Huang2012, Mancinelli2014} and their wide success is  facilitated by the use of powerful and simple mathematical tools such as the transfer matrix method (TMM) \cite{Yariv2000} and temporal coupled-mode theory (TCMT) \cite{Haus1991b,Suh2004, Little1997d}. 
In TMM, the response of a resonant system is calculated directly from the combined interference of light propagating through multiple optical paths and therefore it is suitable to describe systems in which these optical paths are well known, such as optical resonators composed of waveguides or using free-space optics.
In TCMT, on the other hand, the supermodes of a resonant system are calculated from the perturbative coupling of lumped resonators and, in addition to conventional resonators in waveguides and free-space, it appeals to a variety of less conventional electromagnetic resonant structures \cite{Verslegers2012,Karalis2015,Jia2015,Zhen2015}. 
As a perturbative model TCMT is however limited to low-loss weakly coupled systems excited around their resonant frequency (small detuning), while TMM covers a broader range of parameters.
%

When both TMM and TCMT models are possible, the latter often provides a simpler formulation.
For instance, TMM calculations can be cumbersome for resonant structures that allow coupling between counter-propagating modes \cite{Wu2015a, offaxis2013, Souza2015e}, while TCMT can provide a more straightforward approach \cite{ Schmidt2012a,Souza2014c}.
Furthermore the TCMT formalism, based on time differential equations, provides a very simple tool to evaluate dynamic responses of resonators \cite{Yu2014e,Muller2014}.
A time-dependent TMM model is also possible \cite{Sacher2008a} but at the expense of complex calculations when dealing with coupled resonators, which have been increasingly deployed for optical modulation \cite{Xu2009b,Yu2014e,Muller2014}, optical computing \cite{Mesaritakis2013,Vinckier2015} and in the study of dynamic phenomena such as nonlinear oscillations in silicon waveguides \cite{Huang2012, Mancinelli2014}.

An important application of coupled resonators is the generation and control of optical dark states. A resonant state is "dark" when it cannot be excited due to the completely destructive interference of light in the optical path connecting the resonator to external light channels \cite{Chak2007, Benisty2009}.
A slight imbalance in such destructive intereference can lead to a weak effective coupling between the resonator and the external channels, originating a high quality factor (high-Q) resonance instead, or quasi-dark state.
The transition between dark and quasi-dark states have been investigated for several applications including light storage \cite{Scheuer2010}, lasers \cite{Gentry2014h}, optical modulation \cite{Sandhu2012} and wireless energy transfer \cite{Hamam2009}, and TCMT has been often used as the modeling tool \cite{Hamam2009, Sandhu2012, Gentry2014h}.

Here we show that TCMT might however fail to describe coupled resonators presenting quasi-dark states.
We investigate a coupled device composed of three ring resonators and we show that the TCMT model predicts a dark state that is in contrast with experimental observations, TMM and FDTD simulations, which yield a high-Q quasi-dark state instead.
We derive a modified TCMT model that allows the proper excitation of the quasi-dark state and reconciles the different methods.

%
\begin{figure*}[!t]		
\begin{center}
\includegraphics[width=\textwidth]{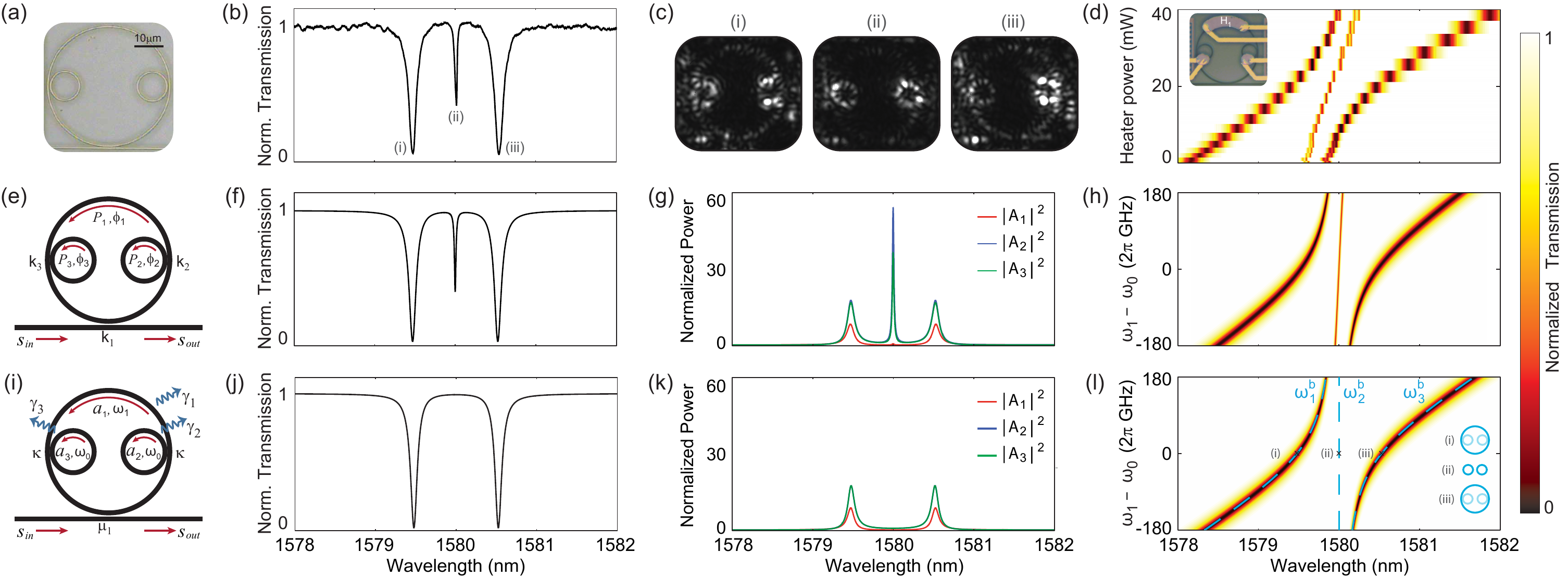}
\caption{
Experimental data (a-d), TMM model (e-h) and TCMT model (i-l) of a three-ring resonator system.
(a) Fabricated device and (b) transmission spectrum showing a triplet with high-Q quasi-dark state in the center when the three rings are degenerate. 
(c) IR-micrograph of the scattered light at each resonance.
(d) Anti-crossing obtained when the outer ring is detuned using a microheater ($H_1$ in the inset micrograph). An overall red-shift is present due to thermal crosstalk affecting the embedded rings.
(e) TMM parameters: $s_{in}$/$s_{out}$ are the input/output fileds; $k_1$ and $k_2$ are coupling coefficients and $\phi_i$ and $P_i$ are the accumulated phase and attenuation in each microring, respectively.
(f) TMM triplet similar to the experimental observation and 
(g) intracavity power spectrum with high power enhancement for the central resonance, in which case light is confined to the embedded rings. $|A_1|^2$, $|A_2|^2$ and $|A_3|^2$ represent the power circulating in the outer ring, first and second embedded rings, respectively. The blue and green curves closely overlap.
(h) TMM anti-crossing showing the evolution of the supermode resonances in the absence of thermal crosstalk.
(i) TCMT model (parameters described in the text). In contrast to the experimental data and TMM, no central resonance is observed in the 
(j) transmission spectrum and 
(k) intracavity power spectrum.
(l) TCMT anti-crossing obtained from the transmission spectrum and from the eigenvalues of $\Omega$ (dashed-blue lines). The central mode is predicted by the eigenvalues but not excited, constituting a dark state in the TCMT model. Inset: spatial distribution of each supermode at degeneracy, representing the eigenvectors of $\Omega$.}
\label{fig1}
\end{center}
\end{figure*}
%
%

\section{Three-ring coupled resonator}
The experimental realization of the three-ring coupled resonator and its spectral response are presented in Fig.\ref{fig1}(a-d).
The device consists of two identical microring resonators coupled to a third dissimilar microring that is coupled to a bus waveguide (Fig.\ref{fig1}(a)).
The transmission spectrum in Fig.\ref{fig1}(b) shows a triplet resulting from the coupling-induced mode-splitting when the three rings are degenerate, while the light distribution in each of the three resonances is illustrated in the infra-red (IR) micrographs of Fig.\ref{fig1}(c).
The lateral resonances of the triplet have simillar Q-factor (14,000) and extinction ratio, whereas the central resonance constitutes a so-called quasi-dark state with significantly higher Q-factor (66,000) as the light is mostly localized in the embedded rings, effectivelly reducing the extrinsic (coupling) losses.
A typical anti-crossing diagram (Fig.\ref{fig1}(d)) is obtained from the transmission spectrum when the detuning between outer and embedded rings is controlled by means of an integrated microheater ($H_1$ in the inset micrograph of Fig.\ref{fig1}(d)). 
The experimental anti-crossing is subjected to an overall red-shift of the resonances due to thermal crosstalk, but the evolution of the supermodes remain clear.
The device was fabricated in a standard silicon-on-insulator (SOI) platform with typical dimensions used for silicon channel waveguides \cite{Barea2013e}:
450-nm by 220-nm waveguides for quasi-TE mode operation with microring radii $R_1$ = 20 $\mu$m, $R_2$ = $R_3$ = 5 $\mu$m, and 200-nm coupling gap between outer and embedded microrings and between the outer microring and the bus waveguide.
The microheaters and contact pads consist of 100-nm Ni-Cr and 2/200-nm Ti/Au films fabricated in a post-process step and the measured electrical resistance of microheater $H_1$ is 130 ohm.

The spectral features obtained experimentally are well reproduced by the TMM model (Fig.\ref{fig1}(e-h))
\dash see \href{link}{Supplement 1} part I for the TMM equations and for numerical values of the parameters used to obtain the plots in Fig.\ref{fig1}. 
The transmission spectrum (Fig.\ref{fig1}(f)) shows a similar triplet with resonance-splitting dictated by the coupling between embedded and outer ring resonators ($k_2$) and with a high-Q central resonance. The intracavity power for each ring (Fig.\ref{fig1}(g)) \dash normalized to $s_{in}$ \dash provides a quantitative assessment of the power distributions observed in the IR-micrographs: for both lateral resonances light circulates in all the three rings, while for the central resonance light is localized within the embedded rings resulting in a small effective coupling to the bus waveguide and high power enhancement.
Finally, the TMM anti-crossing diagram of Fig.\ref{fig1}(h) provides the expected evolution of the supermodes in the absence of thermal crosstalk.

Unlike TMM, however, the TCMT model disagrees with the experimental observations as it predicts a dark state for the central supermode (Fig.\ref{fig1}(i-l)).
No central resonance notch appears in the transmission spectrum (Fig.\ref{fig1}(j)) and no light circulates in the cavity (Fig.\ref{fig1}(k)) since the supermode cannot be excited by the incoming light.
These results are calculated using a general formulation of the orthogonal TCMT \cite{Suh2004} (see \href{link}{Supplement 1}, part II),
\vspace{-5pt}
\begin{align}									
& \frac{d\vec{a}}{dt} = (j \Omega - \Gamma) \cdot \vec{a} + K^{\tr} \cdot \vec s_{in}
\label{cmt_a1}
\\
& \vec s_{out} = C \cdot \vec s_{in} + K \cdot \vec{a} 					
\label{cmt_a2}
\end{align}
with the following parameters for the three-ring model (Fig.\ref{fig1}(i)) \cite{Sandhu2012,Hamam2009}: the mode amplitude of the individual resonators are grouped in the mode vector $\vec{a} = (a_1 \ \ a_2 \ \ a_3)^{\tr}$ and their bare resonance frequencies ($\omega_1$ for the external ring and $\omega_0$ for the identical embedded rings) and mutual coupling ($\kappa$) constitute the system matrix 
\vspace{-5pt}
\begin{equation}			
\Omega=
\left( \begin{array}{ccc}
\omega_1		& \kappa	 	& \kappa   \\
\kappa	 		& \omega_0  	& 0			 \\
\kappa			& 0			 	& \omega_0
\end{array} \right);
\label{M_Omega}
\end{equation}
the single bus waveguide is described by the incoming and outgoing power amplitudes $s_{in}$ and $s_{out}$ and requires $C=1$; 
the coupling between the bus waveguide and the resonant system, occurring only through the outer ring, is represented by the coupling vector
\begin{equation}				
K = 
\left( \begin{array}{ccc}
j \mu_1 & 0 & 0
\end{array} \right)
\label{K0}
\end{equation}
where $\mu_1$ represents the coupling of the first resonator to the bus waveguide;
the decay matrix $\Gamma = \Gamma^{loss}+\Gamma^{port}$ completes the model, accounting for the intrinsic losses in each resonator, $\Gamma^{loss} = \text{diag}(\gamma_1 \ \ \gamma_2 \ \ \gamma_3)$ and for the extrinsic loss term $\Gamma^{port} = \text{diag}(\mu_1^2/2 \ \ \ 0 \ \ \ 0)$ according to eq.(\textbf{S11}).
The transmission spectrum is calculated as $|s_{out}/s_{in}|^2$ while the intracavity power spectra are calculated using eq.(\textbf{S9}).

The prediction of a dark state by TCMT can be understood considering the interaction between the supermodes of the coupled system and the bus waveguide.
First, we calculate the eigenvalues and normalized eigenvectors of $\Omega$, which give the supermodes' resonance frequencies $\omega^b_i$'s and mode amplitudes $b_i$'s. The  $\omega^b_i$'s are depicted in Fig.\ref{fig1}(l) (blue traces) and they follow closely the spectral evolution expected from the experimental results and TMM. At degeneracy, the eigenfrequencies and eigenmodes are
\vspace{-5pt}
\begin{equation}			
\begin{array}{llll}
\omega_1^b = &	\omega_0 + \sqrt{2}\ \kappa & , \ \ \ &
b_1 = \begin{array}{ccc} \Big( \frac{-1}{\sqrt{2}} &  \frac{1}{2} & \ \frac{1}{2} \ \Big)^{\tr} \end{array} \\	

\omega_2^b = &	 \omega_0  & , \ \ \ & 
b_2 = \begin{array}{ccc} \Big(  \  0	&  \frac{-1}{\sqrt{2}} & \frac{1}{\sqrt{2}} \Big)^{\tr}\end{array} \\

\omega_3^b = &	\omega_0 - \sqrt{2}\ \kappa & , \ \ \ &
b_3 = \begin{array}{ccc} \Big(  \frac{1}{\sqrt{2}} &  \frac{1}{2} & \ \frac{1}{2} \ \Big)^{\tr}. \end{array} \\
\end{array}
\label{eq_eigensystem}
\end{equation}
These expressions show that $b_2$, the supermode corresponding to the (quasi-)dark state, is completely confined to the embedded rings while $b_1$ and $b_3$ have components in the outer ring as illustrated in the inset of Fig.\ref{fig1}(l).
Since $b_2$ vanishes in the outer ring it cannot be excited by the incoming light which only couples to $a_1$ (see eq.(\ref{K0})). 
The effective zero drive for supermode $b_2$ can be directly seen rewriting $K$ in the coupled basis,
\begin{equation}
K^b = j \frac{\mu_1}{\sqrt{2}} 
\left( 
\begin{array}{ccc}
-1 & 0 & 1
\end{array} \right)
\label{Kb0}
\end{equation}
calculated as $K^b = K \cdot (S^{-1})^{\tr}$, where $S$ is the similarity matrix formed by the column eigenvectors of eq.(\ref{eq_eigensystem}).
$K^b$ represents the coupling between the input/output power amplitudes ($s_{in}$/$s_{out}$) and supermodes $b_1$, $b_2$ and $b_3$ and it shows that the coupling to $b_2$ is effectively zero.
%

\begin{figure}[!t]		
\begin{center}
\includegraphics[width=\columnwidth]{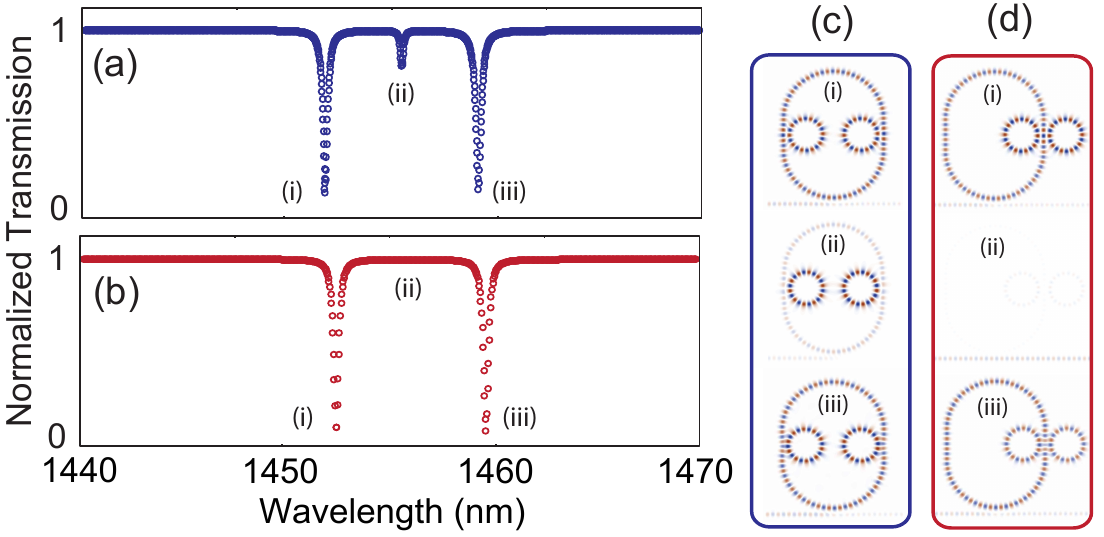}
\caption{2D-FDTD simulations. (a,b) Transmission spectrum and (c,d) steady-state electric field amplitude of the supermodes of the three-ring device at degeneracy in two different configurations. (a,c) When the embedded rings coupled to the outer ring at different positions a weak field circulates in the outer ring allowing the excitation of the quasi-dark state (ii). (b,d) When the embedded rings couple to the outer ring at the same position the destructive interference in the outer ring prevents the excitation of supermode (ii), originating a dark state.}
\label{fig2}
\end{center}
\end{figure}

FDTD simulations of two distinct three-ring designs allow to understand the limitations of the TCMT model and how it can be modified to properly describe the quasi-dark state.
The transmission spectrum and the mode profiles of each supermode are presented in Fig.\ref{fig2}(a,c) for a design similar to the one described in Fig.\ref{fig1} with two embedded rings coupled to the outer ring at different positions, while Fig.\ref{fig2}(b,d) shows these results for a design where both rings are coupled to the outer ring at the same point.
The FDTD simulation for the first design is consistent with the previously discussed experimental and TMM results as it also predicts the excitation of the quasi-dark state.
In the second design, on the other hand, the supermode is not excited and constitutes therefore an effective dark state in agreement with the TCMT prediction.
In a lumped element model such as TCMT, however, these two designs are equivalent: two identical resonators weakly coupled to a third one which in turn is coupled to a bus waveguide, with system matrix and coupling vector given by eq.(\ref{M_Omega},\ref{K0}) and supermodes given by eq.(\ref{eq_eigensystem}).

The fundamental difference between the two designs which not accounted in the TCMT model lies on the effect of the embedded rings in the roundtrip phase of the outer ring.
In the absence of embedded rings the outer ring is resonant at $\omega_0$ and its accumulated roundtrip phase is a multiple of $2 \pi$.
When the two embedded rings couple to the outer ring at different positions, each of them introduces a zero or $\pi$ phase-shift depending on its coupling regime (undercoupled or overcoupled, respectively \cite{Wu2013a}) and the accumulated roundtrip phase in the outer ring remains a multiple of $2 \pi$.
Even though no resonant light circulates in the outer ring as indicated by $b_2$ in eq.(\ref{eq_eigensystem}) an amount of non-resonant light is able to propagate over the outer ring to feed the embedded rings as illustrated in Fig.\ref{fig2}(c-ii).
In other words, the outer ring acts as a waveguide in this situation, allowing the communication between the supermode $b_2$ confined in the embedded rings and the bus waveguide. 
The contribution of this non-resonant light is not considered in the TCMT model, which only accounts for the resonant mode amplitudes. 
On the other hand, when the coupling between rings occurs in the exact same point a $\pi$ phase shift is introduced in the accumulated roundtrip phase of the outer ring, resulting in destructive interference and preventing the excitation of $b_2$, as shown in Fig.\ref{fig2}(d-ii).
Detailed information regarding the parameters used in the FDTD simulations can be found in \href{link}{Supplement 1} part III. Notice that we used a racetrack as the outer resonator to assure the coupling between outer and embedded rings is the same in both designs.

The TCMT model can nonetheless be altered to deliver a description of the three-ring design that allows for the excitation of the quasi-dark state $b_2$.  
This is accomplished with a modified coupling vector which will be derived in the next section.


\section{TCMT with modified coupling vector}
%
In this section, we propose a modified TCMT model that incorporates new terms to the coupling vector $K$ and yields results in agreement with TMM and FDTD.
The coupling vector for the three-ring system can be written in its most general form, according to eqs.(\textbf{S10}, \textbf{S11}), as $K=j(\mu_1 \ \ \mu_2 \ \ \mu_3)$ where $\mu_i \in \mathds{R}$. 
A modification of $K$ requires a modification of $\Gamma^{port}$, whose components are given by $\Gamma^{port}_{i j} = \frac{\mu_i \  \mu_j}{2}$.
Therefore, modifying the TCMT model reduces to deducing the correct expressions for the coupling terms $\mu_i$.
%

We seek to write the $\mu_i$'s in terms of the power coupling coefficients $k_i$'s defined in Fig.\ref{fig1}(e) in order to establish a direct correspondence between TCMT and the power coupling parameters used in TMM.
The first term $\mu_1$ is the usual power-energy coupling coefficient for a bus-ring configuration \cite{Little1997d}, written in terms of $k_1$ as
\begin{equation}	
\mu_1 = k_1 \sqrt{\frac{v_g}{L_1}}
\label{mu_1}
\end{equation} 
where $v_g$ is the group velocity in the outer ring (we will assume the same group velocity for all rings). 
In the coupled basis, the general coupling vector is
\begin{equation}   
K^{b} \equiv
\left(
\begin{array}{c}
K^b_1 \\ K^b_2 \\ K^b_3
\end{array}
\right)^{\tr}
=
\left(
\begin{array}{c}
   -j \frac{1}{2} (\sqrt{2}\ \mu_1 -\mu_2 - \mu_3) \\
   -j \frac{(\mu_2 - \mu_3)}{\sqrt{2}}				  \\
\ \ j \frac{1}{2} (\sqrt{2}\ \mu_1 +\mu_2 + \mu_3) \\
\end{array}
\right)^{\tr}
\label{K_b_gen}
\end{equation}
It is expected that supermodes $b_1$ and $b_3$ be equally coupled to the bus waveguide ($|K^b_1|^2 =|K^b_3|^2$) as the have the same mode profile, thus requiring $\mu_2$ and $\mu_3$ to satisfy $\mu_3 = - \mu_2$. This allows to simplify $K^b$ to
\begin{equation}	
K^{b}= j \frac{1}{\sqrt{2}} \ 
\left( \begin{array}{ccc}
- \mu_1   &
- 2 \ \mu_2	 		&
 \mu_1	
\end{array} \right).
\label{K_b}
\end{equation}
This expression gives the same coupling to $b_1$ and $b_3$ as eq.(\ref{Kb0}) which was already in agreement with the expected results. As for mode $b_2$, it can now be excited by a non-null $\mu_2$.
%

The term $\mu_2$ represents the indirect coupling between bus waveguide and embedded rings and its dependence with $k_1$ and $k_2$ is determined using power conservation \cite{Suh2004, Little1997d}.
Consider a lossless system ($\Gamma^{loss}=0$) with mode $b_2$ excited to energy $|b^0_2|^2$ at \textit{t} = 0.
With no incoming light ($s_{in} = 0$) the energy in the resonator decays and the power flowing through the output port is $ |s_{out}(t)|^2 = 2 \  \mu_2^2 \  |b_2 (t)|^2 $.
The same scenario can be described using a power-normalized amplitude $B_2(t)$ that couples to the outgoing wave $s_{out} (t)$ through a power coupling coefficient defined as $k_{b}$ so that
$|s_{out}(t)|^2 = k^2_b \  |B_2(t)|^2$. The equivalence between the two pictures requires
\begin{equation}	
2 \  \mu_2^2 \  |b_2 (t)|^2
\ = \ 
k^2_b \  |B_2(t)|^2.
\label{Power-energy-B}
\end{equation}
On the other hand, the relation between circulating power and stored energy given by eq.(\textbf{S9}) requires
\begin{equation}	
|B_2(t)|^2 = |b_2(t)|^2 \ \frac{v_g}{2 \  L_2}
\label{B-b}
\end{equation}
where $L_2$ is the length of each identical embedded ring so that $2 \  L_2$ is  the effective length of supermode $b_2$. Eqs.(\ref{Power-energy-B}, \ref{B-b}) allow to write $\mu_2$ in terms of $k_b$:
\begin{equation}	
\mu_2 = \frac{k_b}{2} \sqrt{\frac{v_g}{L_2}}
\label{mu2-kb}
\end{equation}
%

Finally, the power coupling coefficient $k_b$ is given by $\frac{k_1 \  k_2}{2}$, as follows. 
The supermode confined to the embedded rings, with power $|B_2|^2$, is fed by a certain amount of power circulating in the outer ring $|A^\pi_1|^2$ by means of $k_2$, so that
\begin{equation}		
|B_2|^2 = k_2^2 \  |A^\pi_1|^2.
\label{B2_A1}
\end{equation}
At frequency $\omega^b_2$, for which this correction is derived, $|A^\pi_1|^2$ is the circulating power in a microring out-of-resonance and can be estimated using TMM along with the TCMT assumptions of low loss ($P_1 \rightarrow 1$) and weak coupling ($k_1 \ll 1$):
\begin{equation}	
|A_1|^2 = \abs{\frac{j k_1 \  P_1 \ e^{j \phi_1}}{1- t_1 \  P_1 \ e^{j \phi_1}} }^2 |s_{in}|^2
\ \ \ \xrightarrow[k_1 \ll 1, \ P_1 \rightarrow 1 ]{\phi_1 \rightarrow \pi} \ \ \
|A^\pi_1|^2 = \frac{k_1^2}{4} |s_{in}|^2
\label{anti-res}
\end{equation}
Combining eq.(\ref{B2_A1}) and eq.(\ref{anti-res}) and reminding that $k_b$ was defined as the coupling coefficient between $B_2$ and the bus waveguide, we have $k_b = \frac{k_1 \  k_2}{2}$.
%

The modification of the TCMT model is therefore complete, consisting of a new coupling vector $K = j(\mu_1 \ \ \mu_2 \ \ -\mu_2)$ and additional elements $\Gamma^{port}_{ij}=\frac{\mu_i \  \mu_j}{2}$ to the decay matrix, where 
\begin{equation}
\mu_1 = k_1 \sqrt{\frac{v_g}{L_1}} \ , \ \ \ \ \ \ \   
\mu_2 = \frac{k_1 \  k_2}{4} \sqrt{\frac{v_g}{L_2}}.
\label{mu1_2}
\end{equation}
The correct response can be obtained by using these parameters in eq.(\ref{cmt_a1},\ref{cmt_a2}).

\section{Validation of the modified-TCMT model}
\begin{figure}[t]
\begin{center}
\includegraphics[width=\columnwidth]{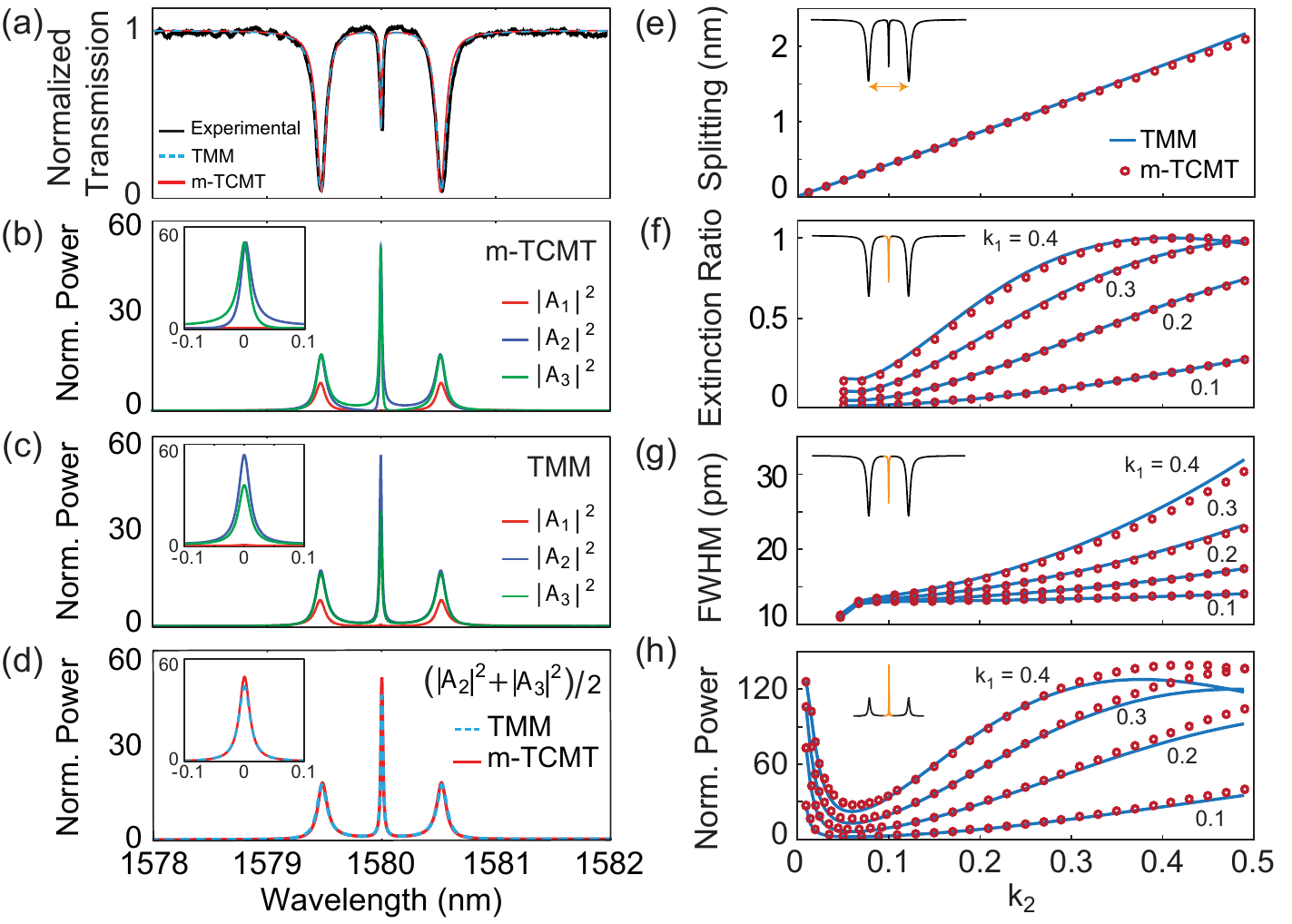}
\caption{
Comparison between modified-TCMT (m-TCMT) and TMM. (a) Transmission spectrum and (b,c) intracavity power spectrum calculated with the parameters used in Fig.\ref{fig1}. Insets: detail of the central peaks.
(d) Average intracavity power for the quasi-dark state calculated with TMM and m-TCMT.
(e) Resonance splitting for the triplet and (f) extinction ration, (g) linewidth and (h) average intracavity power for the quasi-dark state calculated for various coupling coefficients. The m-TCMT calculations agree with TMM for a wide range of coupling strengths, while in the standard TCMT curves (f-h) would vanish.}
\label{fig3}
\end{center}
\end{figure}
%
%

We validate the modified-TCMT (m-TCMT) steady-state solution by its comparison with TMM.
The transmission spectrum of (Fig.\ref{fig3}(a)) shows that m-TCMT closely reproduces the experimental and TMM traces, including a clear high-Q central resonance associated with an excited quasi-dark state.
 Although the intracavity power spectra calculated with m-TCMT (Fig.\ref{fig3}(b)) and with TMM (Fig.\ref{fig3}(c)) reveal some differences for the power circulating in the embedded rings (inset figures), their average power for supermode $b_2$, calculated as $(|A_2|^2 + |A_3|^2)/2$, are in very good agreement (Fig.\ref{fig3}(d)).
The asymmetry between blue and green traces in m-TCMT (inset of Fig.\ref{fig3}(b)) reflects the asymmetry of the cross-decay terms $\Gamma^{port}_{12}$ and $\Gamma^{port}_{13}$ in
\vspace{-5pt}
\begin{equation}
\Gamma^{port}=
\left( \begin{array}{ccc}
\frac{\mu_1^2}{2}  & \frac{\mu_1\  \mu_2}{2}   & \frac{-\mu_1\  \mu_2}{2}   \\
\frac{\mu_1\  \mu_2}{2}	& \frac{\mu_2^2}{2} 	& \frac{-\mu_2^2}{2}	 \\
\frac{-\mu_1\  \mu_2}{2}	&  \frac{-\mu_2^2}{2}	   &\frac{\mu_3^2}{2}
\end{array} \right)
\label{Gamma}
\end{equation}
due to the fact that $\mu_3 = - \mu_2$ in our model.
On the other hand, the power imbalance between  $|A_2|^2$ and $|A_3|^2$ in TMM (inset of Fig.\ref{fig3}(c)) reflects the fact that light arrives at the second embedded ring modified by the resonance of the first one, as can be seen in the expression
\vspace{-5pt}
\begin{equation}
\frac{|A_3|^2}{|A_2|^2} = P_1 \ |\chi_2|^2
\label{A32}
\end{equation}
where $P_1$ is the roundtrip attenuation factor of the outer ring and $\chi_2$ is the complex transmission of the first embedded ring (see \href{link}{Supplement 1} part I).

The new terms in the m-TCMT equations depend on the coupling coefficients $k_1$ and $k_2$, therefore the model must be validated over a wide range of these parameters.
The m-TCMT and TMM models yield very close predictions for various values of coupling coefficients as demonstrated in Fig.\ref{fig3}(e-h), which shows results for the resonance splitting (Fig.\ref{fig3}(e)), the extinction ratio and linewidth of the quasi-dark state resonance (Fig.\ref{fig3}(f,g)) and its average intracavity power (Fig.\ref{fig3}(h)).
Particularly, the agreement between TMM and m-TCMT for the average intracavity power indicates that the
power imbalance captured in the TMM model do not significantly affect the total power in the supermode.  
The two models show slight discrepancies only for combinations of large coupling strengths, when the weak-coupling assumption of TCMT starts to fail.
Notice that, except for the resonance splitting, the calculated quantities would vanish in the standard TCMT model.
A similar comparison for the lateral resonances of the triplet (supermodes $b_1$ and $b_3$) is unnecessary as they were already well described by the standard TCMT and their steady-state values are not affected by $\mu_2$, as predicted by eq.(\ref{K_b}).
The results presented in Fig.\ref{fig3}(e-h) where calculated assuming the same ring radii and effective index used in previous plots and presented in \href{link}{Supplement 1} part I, whilst the attenuation parameters where $P_1 = 0.992$ and $P_1 = 0.998$ (equivalent to $\alpha_1 = \alpha_2 = 5$ dB/cm). 

%
\begin{figure}[t]
\begin{center}
\includegraphics[width=\columnwidth]{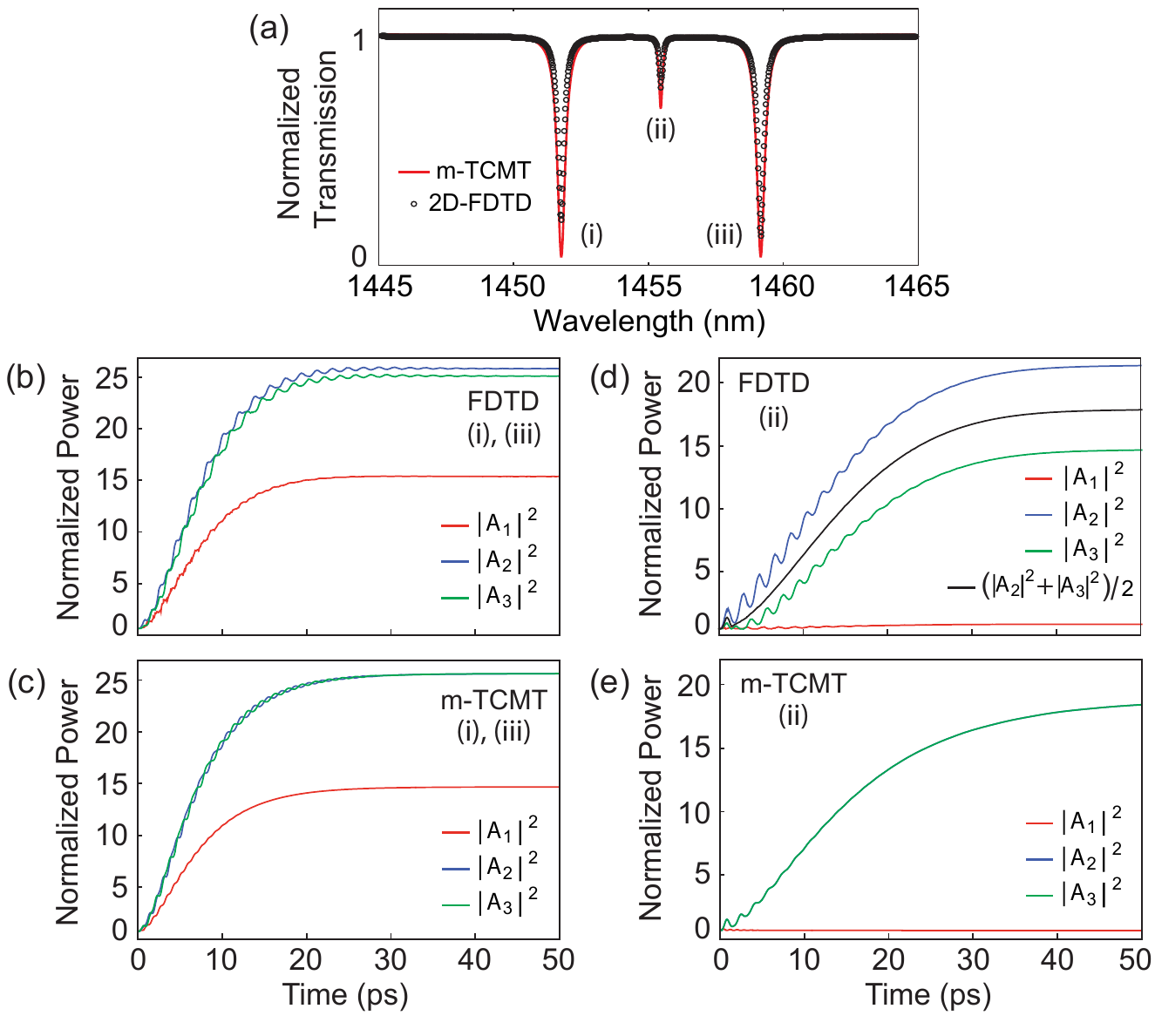}
\caption{Comparison between m-TCMT and 2D-FDTD simulations. (a) Transmission spectrum and (b-e) transient intracavity power evolution. The FDTD results are presented in (b) for the lateral resonances and in (d) for the quasi-dark state, while the corresponding m-TCMT solutions are presented in (c) and (e). The m-TCMT model reproduces the transient evolution for the lateral modes including the fast oscillations presented in the FDTD simulation. For the quasi-dark state it describes the average power circulating in the resonator.}
\label{fig4}
\end{center}
\end{figure}
%
%

In addition to the steady-state response, the m-TCMT model provides a good description of the transient behavior of the coupled system as confirmed by a comparison with 2D-FDTD simulations (Fig.\ref{fig4}) \dash see \href{link}{Supplement 1} part III for simulation parameters.
The transient evolution of the lateral resonances (Fig.\ref{fig4}(b)) presents fast oscillations determined by the resonance splitting ($2\sqrt{2} \kappa$) which are also reproduced by m-TCMT (Fig.\ref{fig4}(c)) as a result of the additional non-diagonal terms in eq.(\ref{Gamma}).
For the quasi-dark state, the FDTD transient presents a power imbalance between embedded rings similar to that observed in the TMM calculations (Fig.\ref{fig4}(d). Once again, it is the average power in the supermode that corresponds to the m-TCMT solution (Fig.\ref{fig4}(e)).
This good description of the transient response of the optical supermodes makes m-TCMT a suitable model to describe dynamic perturbations on the coupled-ring system such as optical modulation through refractive index perturbation.

\section{Conclusion}
In this article we demonstrated an important limitation of the TCMT model: the incorrect prediction of a dark-states in coupled resonators instead of the actual high-Q quasi-dark states.
Through the analysis of a three-ring resonator system, we showed this inaccurate prediction occurs due to the inability of TCMT to account for non-resonant light circulating in the system.
The existence of such non-resonant light can be properly accounted for by introducing extra drive terms in a modified-TCMT (m-TCMT) model.

The TCMT formalism is applied to many fields and the limitation demonstrated here might be present in distinct resonant structures.
We believe our discussion will be helpful to prevent the misrepresentation of quasi-dark states and will provide insight on how the standard TCMT model can be modified to allow accurate results.

\section*{Funding Information}
Conselho Nacional de Desenvolvimento Científico e Tecnológico (National Council for Scientific and Technological Development) (08/57857-2, 156468/2015-8); Fundação de Amparo à Pesquisa do Estado de São Paulo (São Paulo Research Foundation) (2012/17765-7, 2014/04748-2).

\section*{Acknowledgments}
The authors acknowledges Felipe Vallini and Felipe Santos for fruitful discussions and CCS-UNICAMP for providing the micro-fabrication infrastructure.

\bigskip \noindent See \href{link}{Supplement 1} for supporting content.

\bibliography{TCMT_manuscript_2}
\end{document}